\documentstyle[twoside,fleqn,espcrc2,epsfig,rotating]{article}

\newcommand{\comment}[1]{}

\newcommand{\bildchen}[3]{
\begin{center}                                        %
\begin{flushleft}                                     %
\makebox{\Large $\displaystyle #2$}                   %
\end{flushleft}                                       %
\mbox{{\epsfig{figure=#1,width=8.cm,
bbllx=1.8cm,bblly=9.2cm,bburx=20.cm,bbury=19.cm}}}   %
\end{center}                                          %
\begin{flushright}                                    %
   {\Large $\displaystyle #3$ \hspace*{5ex}}          %
\end{flushright}}                                     %
\title{
{\begin{flushright}
       {\tt \small hep-ph/9611329\\
        HUTP-96/A052\\
        TTP96-20\\
	November 1996}
       \end{flushright}}
Structure Functions in Semihadronic Tau Decays}

\author{Gilberto Colangelo\address{INFN - Laboratori Nazionali di Frascati,
Gruppo Teorico, P.O. Box 13, I-00044 Frascati (ROMA), Italy},
Markus Finkemeier\address{Lyman Laboratory of Physics,
        Harvard University,
        Cambridge, MA 02138, USA}%
	\thanks{Talk presented by Markus Finkemeier at the Fourth International Workshp on Tau Lepton Physics (TAU 96), 16--19 September 1996, Estes Park,
Colorado, USA},
Erwin Mirkes and Res Urech\address{Institut f\"ur Theoretische Teilchenphysik,
Universit\"at Karlsruhe, D-76128 Karlsruhe, Germany}}
\begin{document}

\begin{abstract}
We review a variety of topics related to hadronic structure
functions in exclusive semihadronic tau decays.
We introduce the concept of structure functions and summarize the
most important concepts. 
We then calculate the decay $\tau \to 3 \pi \nu_\tau$ for very small
hadronic invariant mass to one loop 
in Chiral Perturbation Theory.
New interesting features emerge with respect to the known results at tree
level, in particular for the structure functions $w_D$ and $w_E$.
We discuss the prospects for experimental verification of our predictions.
Finally, we discuss various issues at higher $Q^2$, related to hadronic 
resonance physics. Here we consider $2\pi$, $\pi K$, $3 \pi$,
$K \pi K$ and $\pi K \pi$ hadronic final states.
\end{abstract}

\maketitle

\section{Overview}
In this note we will present a variety of important
topics related to differential distributions in semihadronic $\tau$ decays.
First, in Sec.~2, 
we will introduce the concept of structure functions as a convenient
tool to extract the information encoded in angular and invariant mass 
distributions.
Then in Sec.~3, 
we will calculate the decay $\tau \to 3 \pi \nu_\tau$
to one loop in Chiral Perturbation Theory.
Finally, in Sec.~4, we will discuss various issues at higher $Q^2$ (resonance
physics), considering $2\pi$, $\pi K$, $3 \pi$, $K \pi K$ and $\pi K \pi$ 
final states.
\section{Hadronic Structure Functions}
Consider two and three meson decay modes of the $\tau$,
\begin{equation}
   \tau \to \left\{ 
   \begin{array}{c} h_1(p_1) h_2(p_2) \\ 
    h_1(p_1) h_2(p_2) h_3(p_3) \end{array}
    \right\} + \nu_\tau
\end{equation}
The amplitudes for these decays can be written as
\begin{equation}
   {\cal M} = V_{\rm CKM} \frac{G_F}{\sqrt{2}} L_\mu H^\mu
\end{equation}
where
\begin{equation}
   L_\mu = \overline{u}_\nu \gamma_\mu \gamma_- u_\tau
\end{equation}
and
\begin{equation}
   H^\mu = \left\{ \begin{array}{r}
   \langle h_1 h_2 | V^\mu - A^\mu | 0 \rangle \\
   \langle h_1 h_2 h_3 | V^\mu - A^\mu | 0 \rangle 
   \end{array} \right.
\end{equation}
The amplitudes can be parametrized by introducing form factors. 
In the general two meson case, there are two such form factors,
\begin{eqnarray}
  \langle h_1 h_2 | V^\mu - A^\mu | 0 \rangle 
  & = &\underbrace{F_V(Q^2) T^{\mu\nu} (p_1 - p_2)_\nu}_{J^P = 1^-}
\nonumber \\ &    
+ &\underbrace{F_S(Q^2) Q^\mu}_{J^P = O^+}
\end{eqnarray}
In the general three meson case, there are four of them,
\begin{eqnarray}
 \lefteqn{\langle h_1 h_2 h_3 | V^\mu - A^\mu | 0 \rangle
     }
\nonumber \\ &= &
\underbrace{[ (p_1 - p_3)_\nu F_1
   +  (p_2 - p_3)_\nu F_2] T^{\mu\nu}}_{J^P = 1^+}
\nonumber \\
   & + & \underbrace{i \epsilon^{\mu\alpha\beta\gamma} p_{1 \alpha} p_{2 \beta}
         p_{3 \gamma} F_3}_{J^P = 1^-}
\nonumber \\
   & + & \underbrace{ Q^\mu F_4}_{J^P = 0^-}
\end{eqnarray}
where the $F_i$ depend on three invariants, e.g.\ $F_i = F_i(Q^2,s_1,s_2)$.
We have used
\begin{eqnarray}
   Q^\mu & := & \left\{ \begin{array}{l} p_1^\mu + p_2^\mu \\ 
    p_1^\mu + p_2^\mu + p_3^\mu
       \end{array} \right.
\nonumber \\
   s_1 & := &(p_2 + p_3)^2 \quad \mbox{and cyclic}
\nonumber \\
  T^{\mu\nu} & := & g^{\mu \nu} - \frac{Q^\mu Q^\nu}{Q^2}
\end{eqnarray}
In specific cases, there are various simplifications. If the two mesons are 
two pions, $h_1 h_2 = \pi^- \pi^0$, then the vector current is conserved
and the scalar form factor vanishes,
$F_S \equiv 0$ for $m_u = m_d$.
In the three pion case, $h_1 h_2 h_3 = \pi^- \pi^- \pi^+$ or
$\pi^0 \pi^0 \pi^-$, Bose symmetry 
relates $F_2$ to $F_1$, via $F_2(Q^2,s_1,s_2) = F_1(Q^2,s_2,s_1)$.
$G$ parity conservation requires $F_3 \equiv 0$ for $m_u = m_d$,
and PCAC requires $F_4 \equiv 0$ for $m_u = m_d = 0$.

So we have seen the the hadronic matrix element for two (three) mesons
in the final state is characterized by two (four) complex functions
of one (three) hadronic invariants.
For experimental analyses, it turns out to be very useful to trade
the two (four) complex functions for four (sixteen) real valued
``structure functions'' $W_X$, which are defined from the hadronic tensor
$H^{\mu\nu}$ in the hadronic rest frame \cite{Kue92},
\begin{equation}
   H^{\mu\nu} := H^\mu H^{\nu\star}
\end{equation}
For the precise definitions, the reader is referred to \cite{Kue92},
the main points are summarized in Tab.~\ref{tabsf}.
The differential decay rate is given by \cite{Kue92}:
\begin{equation}
   d \Gamma =
   \frac{G_F^2}{4 M_\tau} |V_{\rm CKM}|^2  
   \sum_{X} L_X W_X \; 
   \,dPS^{(4)}
\end{equation}
Note that $W_A$, $W_B$ and $W_{SA}$ alone determine $d\Gamma / dQ^2$,
\begin{eqnarray}
\lefteqn{\frac{d\Gamma(\tau\to h_1 h_2 h_3 \nu_\tau}{dQ^2} \propto
   \frac{(M_\tau^2 - Q^2)^2}{Q^4} }
\nonumber \\
& \times &  \int ds_1\, ds_2\, 
  \left\{ (1 + \frac{Q^2}{M_\tau^2})  \frac{W_A + W_B}{6} +
  \frac{W_{SA}}{2} \right\}.
\nonumber \\ 
\end{eqnarray}
(Almost) all structure functions can be determined and disentangled
from each other by studying angular correlations of the hadronic system,
for details see \cite{Kue92}.
\begin{table}
\caption{The structure functions} \label{tabsf}
$$
\begin{array}{|c|c|c|c|}
\hline 
& \multicolumn{3}{|c|}{H^\mu \longrightarrow}\\
H^{\nu\star} \downarrow   & J^P = 1^+ & J^P = 1^- & J=0 \\
\hline 
 & & & \\
J^P = 1^+ & 
{\bf W_A}
& & \\
 &  W_C W_D W_E & & \\
 & & & \\
\hline
 & & & \\
J^P = 1^- & W_F W_G & {\bf W_B} &  \\
& W_H W_I & & \\
 & & & \\
\hline
 & & & \\
J = 0     & W_{SB}  W_{SC} 
& W_{SF} W_{SG} &
     {\bf W_{SA}}\\
 & W_{SD} W_{SG} & & \\
 & & & \\
\hline \hline
& & \multicolumn{2}{|c|}{\underbrace{\hspace*{15ex}}_%
{\displaystyle h_1 h_2}}\\
\hline
& \multicolumn{3}{|c|}{\underbrace{\hspace*{30ex}}_%
{\displaystyle h_1 h_2 h_3}}\\
\hline
\end{array}
$$
\end{table}

\section{Low Energy Expansion of the Decay $\tau \to 3 \pi \nu_\tau$}
{\bf 1.} 
Tau decays into three pions involve hadronic invariant masses $\sqrt{Q^2}$
from threshold $3 M_\pi$ up to $M_\tau$. Therefore theoretical predictions
are difficult, and one has to resort to phenomenological models. 
However, for $\sqrt{Q^2}$ below about $600\,\mbox{MeV}$, chiral perturbation
theory (CHPT) \cite{gasser84} 
allows to calculate the hadronic matrix elements in a systematic 
expansion in external momenta and quark masses.

This is interesting for two reasons. First, this calculation allows to 
understand the complicated three pion dynamics systematically
in a small corner of phase space, which can be used
to test and improve phenomenological models.
Second, the small $Q^2$ region could become accessible experimentally
in future at $b$ and $\tau$-charm factories.

Here we want to emphasize a few new and interesting features
arising from the one loop calculation in CHPT. We will explain their origin in
detail, and claim that they should be seen by the new high statistics
experiments. 
\\[5mm]     
%
{\bf 2.} 
Sometimes it is  useful to parametrize the hadronic 
matrix element using conventions which differ from 
the notation introduced in Sec.~2. 
One can also use three different form factors $F$, $G$ and $H$, and
introduce isospin indices:
\begin{eqnarray}
   H_\mu &=& \langle \pi^i(p_1) \pi^j(p_2) \pi^k(p_3) | A^l_\mu(0) | 0 
   \rangle \nonumber \\
  &=& \delta^{ij}\delta^{kl}\Big[(p_1 + p_2)_\mu G(s_1,s_2,s_3) 
\nonumber \\ &+ & 
          (p_1 - p_2)_\mu H(s_1,s_2,s_3)  
\nonumber \\ & + &  
          p_{3\,\mu} F(s_1,s_2,s_3) \Big] 
   + \mbox{permutations}
\end{eqnarray}
$F$, $G$ and $H$ are isospin and Lorentz scalars and can be
written as functions of
$s_1$, $s_2$ and $s_3$.
This general matrix element can be projected on a
physical matrix element in a trivial manner: isospin symmetry allows
to relate the matrix element of the different decay channels.
We have calculated them in the framework of CHPT to one loop. The
complete analytic results and a very detailed analysis are given
elsewhere \cite{long}. 
Here we would like to discuss some interesting features of this decay
as predicted by CHPT at the level of structure functions, showing how
they derive from basic properties of the form factors.

We will
consider only four structure functions $W_X$, 
the ones related only to the spin 1 part
of the matrix element. They can be expressed as:
\begin{eqnarray}
\label{Ws}
W_A &=&(x_1+x_2)^2|K|^2
\nonumber \\ & + & \left[(x_1-x_2)^2+4x_3^2\right] |H|^2 
\nonumber \\ &+ &
  2(x_1^2-x_2^2) \mbox{Re}(K H^*) \nonumber \\
W_C &=& W_A-8 x_3^2 |H|^2 \nonumber \\
W_D &=& 4 x_3\Big[(x_1-x_2)|H|^2
\nonumber \\ & +& (x_1+x_2)\mbox{Re}(K H^*)\Big] 
\nonumber
\\
W_E &=& 4 x_3 (x_1+x_2) \mbox{Im}(K H^*) \; ,
\end{eqnarray}
where $x_i$ are simple kinematical functions (see Ref. \cite{Kue92}
for details), and $K=(G-F)/3$.
$W_A$ governs the rate and the distributions in the Dalitz plot, 
while the remaining functions determine the angular distribution.

{}From CHPT to one loop we find:

\begin{eqnarray}
K &=& \frac{i}{F_\pi} \Big\{ \frac{2}{3} + \frac{1}{3 F_\pi^2}
\Big[s_3\bar{J}(s_3)-\frac{M_\pi^2}{2} 
\Big(\bar{J}(s_1)
\nonumber \\ & + & {}
\bar{J}(s_2) \Big) \Big] 
 + \frac{1}{288\pi^2 F_\pi^2} \Big[ 4 \bar{l_1}(s_3-2 M_\pi^2)
\Big. \Big. \nonumber \\ 
&-& \Big. 2 \bar{l_2}(s_1+s_2-4M_\pi^2) 
+ \Big. 12M_\pi^2 \bar{l_4} +
2\bar{l_6}Q^2
\nonumber \\ & + & {}
s_1+s_2 -6(s_3+M_\pi^2) \Big] \frac{}{}+O(p^4)\Big\}
\nonumber \\ 
H &=& \frac{i}{F_\pi} \Big\{\frac{1}{6 F_\pi^2}\Big[
\Big(5M_\pi^2-2s_1 \Big) \bar{J}(s_1) 
\nonumber \\ & -& 
\Big(5M_\pi^2-2s_2 \Big) \bar{J}(s_2) \Big] \Big. \nonumber \\
&+& \Big. \frac{1}{96 \pi^2 F_\pi^2} \Big(\frac{5}{3}-2\bar{l_2}
\Big)(s_1-s_2) 
\nonumber \\ & + & O(p^4)  \Big\}\; .
\label{ff}
\end{eqnarray}
where
\begin{eqnarray}
   \bar{J}(s) & = & \frac{1}{16 \pi^2}\left(\sigma \ln 
                 \frac{\sigma - 1}{\sigma + 1} + 2 \right)
\nonumber\\
\sigma     & = & \sqrt{1 - 4 M^2_\pi /s}
\end{eqnarray}
The $\bar{l}_i$ are the renormalized coupling constants of CHPT
at $O(p^4)$ (see \cite{long} for numerical values).

As required by Bose symmetry $K$ ($H$) is symmetric
(antisymmetric) under the exchange of $s_1$ and $s_2$. 
Isospin symmetry requires that the form factors in the $2 \pi^0 \pi^-$ 
mode are equal to the isospin symmetric form factors (\ref{ff})
multiplied by a factor $\sqrt{2}$, whereas the form factors in the $2
\pi^- \pi^+$ mode are given by:
\begin{eqnarray}
K^{(--+)} &=&-\frac{1}{\sqrt{2}}\Big[ K(s_1,s_3,s_2)+K(s_2,s_3,s_1)
\nonumber \\ &-& 
H(s_1,s_3,s_2)-H(s_2,s_3,s_1) \Big] \nonumber \\
H^{(--+)} &=& \frac{1}{\sqrt{2}} \Big[3(
K(s_1,s_3,s_2)-K(s_2,s_3,s_1) ) 
\nonumber \\
&+& H(s_1,s_3,s_2)-H(s_2,s_3,s_1) \Big] \nonumber \; .
\end{eqnarray}

{\bf 3.} Numerical results for the structure functions
are given in Figs. 1--3.
We plot $s_1$, $s_2$ averaged functions $w_X(Q^2) = $
$$
 = \left\{
   \begin{array}{l}
      \int ds_1 \, ds_2 W_X(Q^2,s_1,s_2) 
    \\ \quad \mbox{for} \quad X=A,C \\ \\
      \int ds_1 \, ds_2 \, \mbox{sign}(s_1-s_2)\, W_X(Q^2,s_1,s_2) 
    \\ \quad \mbox{for} \quad X=D,E
   \end{array}
   \right.
$$
where $Q^\mu = p_1^\mu + p_2^\mu + p_3^\mu$.

Corrections to the tree
level results soon become very important. Because of this, and because
of the obvious difficulty in accessing experimentally the region close
to threshold,  
significant direct tests of 
these predictions in the low energy region will
require very high statistics.
The $O(p^4)$ chiral prediction for 
the branching ratio for $\Big(\tau \to \nu_\tau +
(3\pi)(Q^2 \leq Q_{\max}^2) \Big)$ is of the order of
$10^{-5}$ for $Q_{\max} =
600\, \mbox{MeV}$ and $10^{-4}$ for $Q_{\max} =
700\, \mbox{MeV}$ \cite{long}. 
This shows that experimental tests in the low energy region
are difficult but not hopeless.
Moreover, some of the qualitative features of the structure
functions, which the chiral
expansion allows to understand in detail,
will also have 
consequences at higher 
energies and so can be tested experimentally more easily.
  
\begin{figure}                                                  
\caption{Integrated structure function $w_A(Q^2)$:
CHPT prediction at tree level (dashed), one loop 
(solid) and from a vector meson dominance model (dotted).
The four functions, $w_A$ and $w_C$ for both modes $2\pi^- \pi^+$
and $2 \pi^0 \pi^-$ all look identical within the resolution of this
diagram.}
\label{figwa}
\bildchen{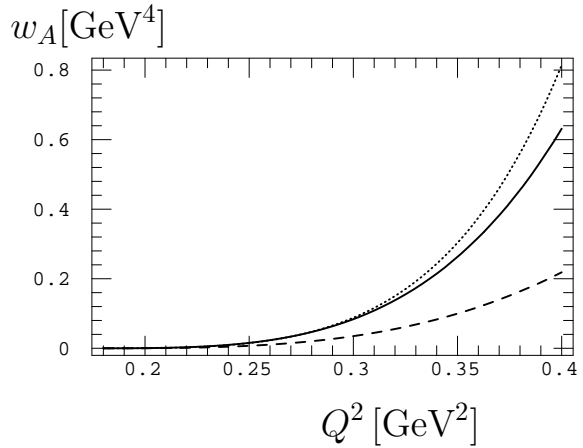}{w_A [\mbox{GeV}^{4}]}%
{{Q^2}\,[\mbox{GeV}^2]}
\end{figure}
The first feature, which the chiral expansion predicts, is that 
 $w_A^{(--+)} \approx
 w_C^{(--+)} \approx
  w_A^{(00-)} \approx
 w_C^{(00-)}$.
In fact, the
four plots can not be distinguished from each other within the
resolution of Fig.~\ref{figwa}.
The equality of the structure functions for the two charge modes indicates
that $w_A$ and $w_C$ are dominated by the [210] partition \cite{pais60}.
We have explicitly verified that this is the case.
The near equality of $w_A$ and $w_C$ at low energies
can be understood algebraically. In Eq. (\ref{Ws}) we read that
\begin{equation}
W_A - W_C =  8 x_3^2 |H|^2 \; .
\end{equation}
$H$ vanishes at tree level according to (\ref{ff}). 
Thus the difference starts as
the {\em square} of a quantity of $O(p^2)$. 
Furthermore, $H$ is kinematically suppressed at low energies, 
because it vanishes for $s_1 = s_2$.
Therefore $w_A \approx w_C$  should be valid even at
energies well above 
those where one would trust a one loop CHPT calculation.

In Fig.~\ref{figwa} we also compare to the prediction from the vector meson
dominance model (VMD) in \cite{kuehn90}. Below $\sqrt{Q^2}$ of about
$600\;\mbox{MeV}$,
where we would trust CHPT, we find very good agreement between 
the VMD model and the one loop chiral prediction.

\begin{figure}           
\caption{Integrated structure function $w_D(Q^2)$. 
CHPT at one loop for $2\pi^-\pi^+$ (solid) and for $2\pi^0\pi^-$
(dashed-dotted) and from the VMD model in \protect \cite{kuehn90} 
(dotted, identical prediction
for both charge modes)}  
\label{figwd}
\bildchen{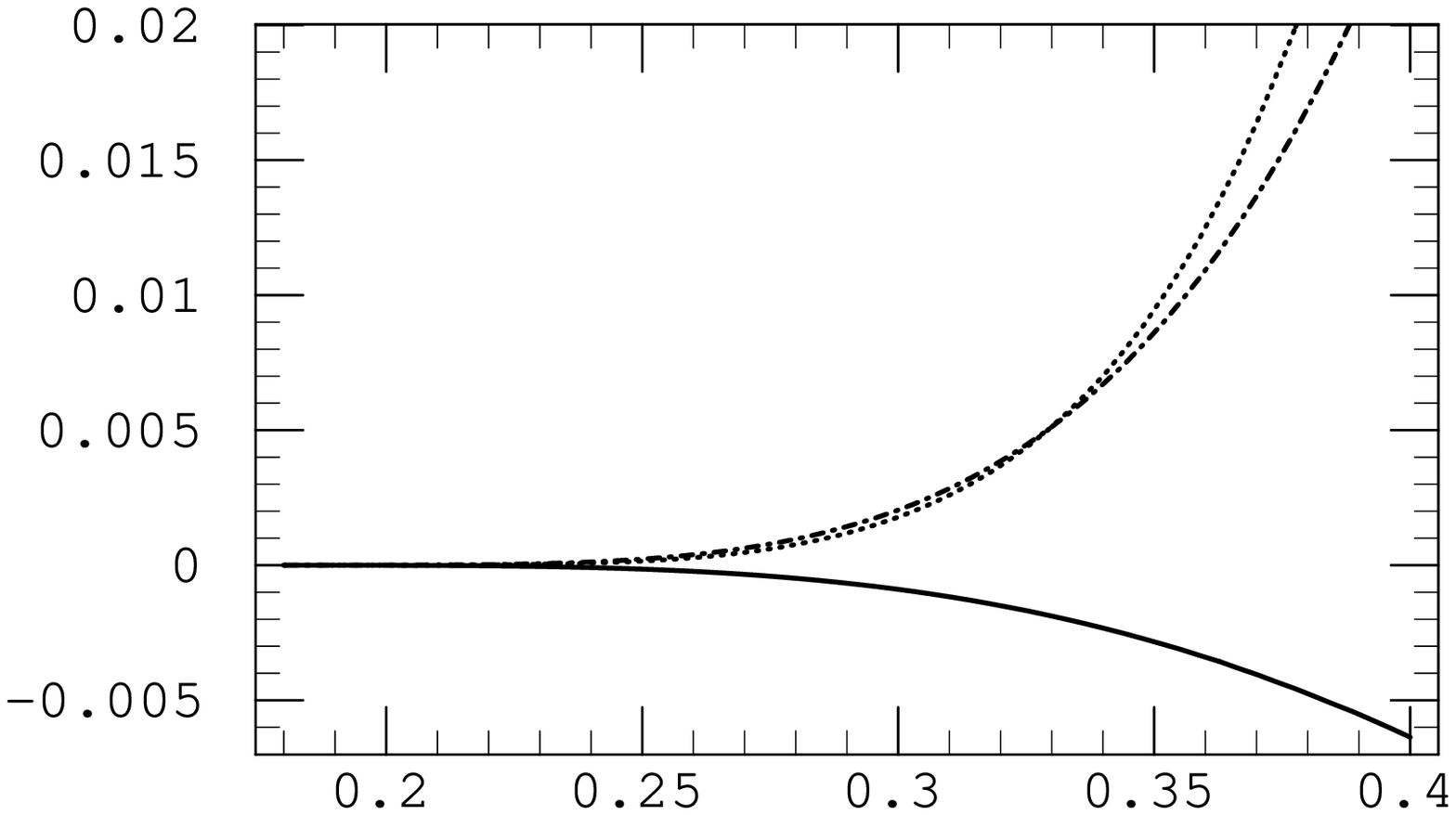}{w_D [\mbox{GeV}^{4}]}%
{{Q^2}\,[\mbox{GeV}^2]}
\end{figure}
\begin{figure}                                                  
\caption{Integrated structure function $w_E(Q^2)$. 
CHPT at one loop for $2\pi^-\pi^+$ (solid) and for $2\pi^0\pi^-$
(dashed-dotted) and from the VMD model in \protect \cite{kuehn90} 
(dotted, identical prediction for both charge modes) }
\label{figwe} 
\bildchen{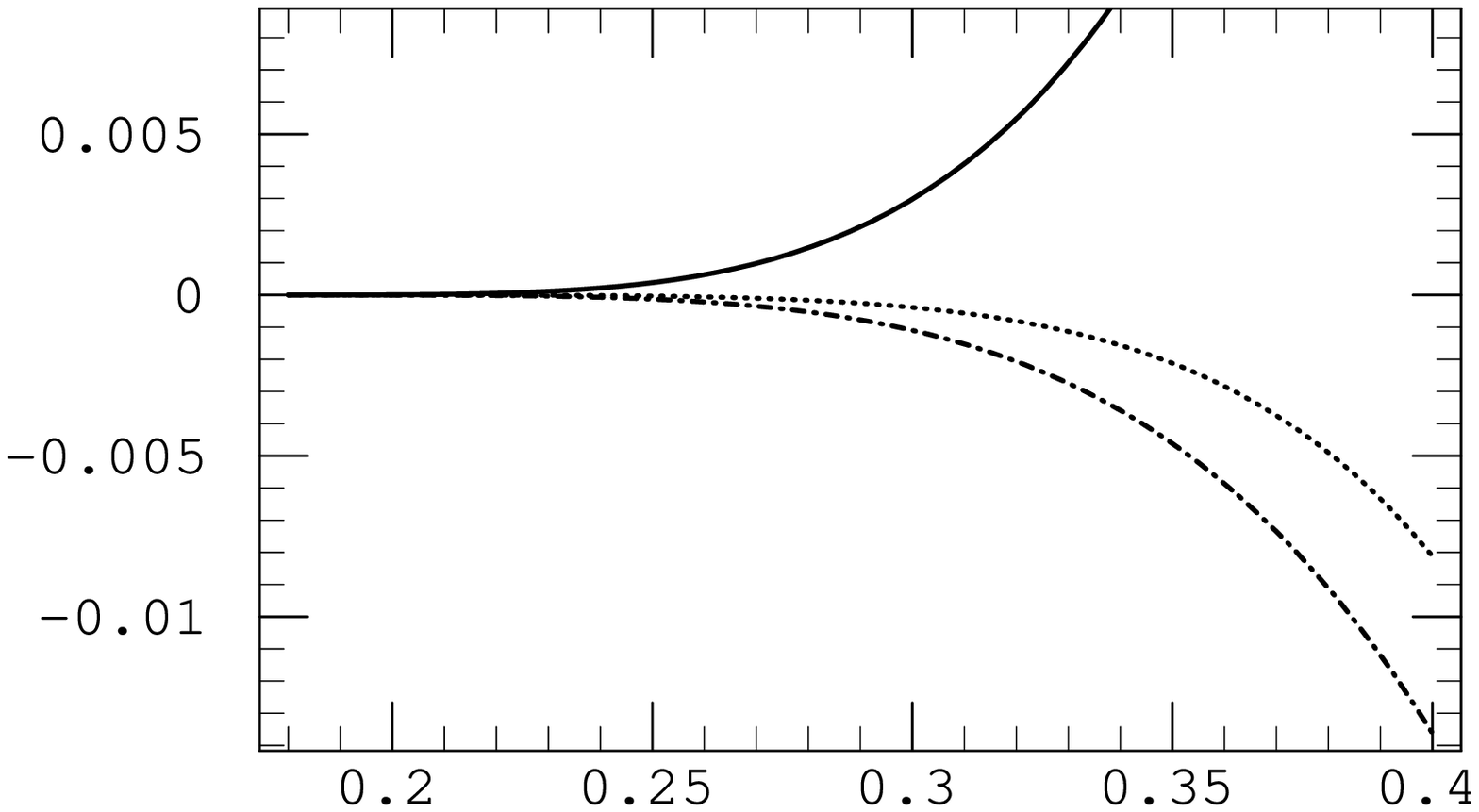}{w_E [\mbox{GeV}^{4}]}%
{{Q^2}\,[\mbox{GeV}^2]}
\end{figure}
{\bf 4.} 
In Fig.~\ref{figwd} and \ref{figwe} we plot $w_D(Q^2)$ and $w_E(Q^2)$.
These are much smaller than $w_A$,
and 
correspondingly more difficult to be measured. 
A measurement of them in the low energy region 
can only be considered at a $b$ or $\tau$-charm factory.

The behaviour of these structure functions is rather interesting. 
The one loop predictions from CHPT differ strongly for the
two charge modes. 
The origin of this sign difference can again be understood in detail:
neglecting $|H|^2$ that as discussed above is tiny near threshold,
$W_D$ and $W_E$ are the real and imaginary part of the same function:
\begin{equation}
4 x_3 (x_1+x_2) K H^*
\end{equation}
The leading contribution to this function comes from the interference
of the tree level part of $K$ and the one loop part of $H$. At tree
level $K$ exhibits a simple sign difference for the two charge modes:
\begin{eqnarray}
K^{(--+)} & = &\frac{i \sqrt{2}}{F_\pi}( - 2/3 + O(p^2))
\nonumber \\
K^{(00-)} & = &\frac{i \sqrt{2}}{F_\pi}( 2/3 + O(p^2))  .
\end{eqnarray}
Since $H$ does not change sign between the two charge modes the same
sign difference as in $K$ at tree level is reproduced in the two
structure functions $W_D$ and $W_E$.
In our numerical calculation the sign difference remains up to
$\sqrt{Q^2} \simeq 0.6 \;\mbox{GeV}$, where large corrections to this
calculation are already expected.

Note that CHPT predicts
$w_D^{(--+)} < 0$ and $w_E^{(--+)} > 0$, 
which should be correct near threshold. 
However, experimental data are available for $Q^2 \geq 0.8\;{\rm GeV}^2$
\cite{argus90}, which indicate 
$w_D^{(--+)} > 0$ and $w_E^{(--+)} < 0$. 
Thus, unless the higher orders in the chiral expansion completely 
change the CHPT result, somewhere between threshold and $Q^2 \sim 0.8\;
{\rm GeV}^2$ there must be a zero for both structure functions in the
$2 \pi^- \pi^+$ mode.  
This feature is absent in  all the phenomenological
models in the literature of which we are aware.
We can not predict from the chiral expansion, whether these zeros will
occur at very low energies or only a little below
$Q^2 = (M_\rho + M_\pi)^2 \approx 0.8\ \mbox{GeV}^2$.
We suggest a careful
experimental examination of $w_D$ and $w_E$ from
$Q^2 = 0.8\ \mbox{GeV}^2$ downwards.

%
{\bf 5.} Let us summarize our main results.
First, we predict that $w_A$ and $w_C$ are very similar at threshold.
This has already been noted in phenomenological models, but we have
shown that $w_A = w_C$ is actually a low energy theorem of QCD, which 
receives corrections only at next-to-next-to leading order.
So the symmetry structure of QCD requires $w_A \approx w_C$ 
well beyond the very low energy region.

Second, near threshold we find a sign difference between the two
charge modes for $w_D$ and $w_E$. Taking into account also the
available experimental data in $2 \pi^- \pi^+$ channel we conclude
that in this channel both structure functions should have a zero somewhere
between threshold and $Q^2\sim 0.8\;{\rm GeV}^2$.
This feature is absent in any of the phenomenological models of which
we are aware, and its experimental verification would be the first
evidence for the presence of the [300] partition state \cite{pais60} 
in tau decays into three pions. 

With the same method we can calculate also the scalar part of the
matrix element close to threshold. It turns out that, compared to the 
one loop prediction from CHPT, this scalar part is 
underestimated, if present at all, in phenomenological models.
A detailed analysis of this and other topics, such as the application of
CHPT to decays into two pions, is given elsewhere \cite{long}.

\section{Issues at Higher $Q^2$: Resonances}
{\bf 1.} {$\pi^- \pi^0$ final states.}
There are some discrepancies between experimental data and predictions
using the ``chirally normalized vector meson dominance'', or
``CN-VMD'' model for $\pi K \pi$ final states \cite{Fin96,erwin}.
Prompted by this, we have started to systematically {\em test} the
assumptions of the CN-VMD model, beginning with the simplest mode, i.e.\
the two pion final state.

A completely general ansatz for the relevant form factor $F_V(Q^2)$ is
\begin{eqnarray}
  F_V(Q^2) & = &
   \frac{f_\rho(Q^2) g_{\rho \pi\pi}(Q^2)}{M_\rho^2} \mbox{BW}_\rho(Q^2)
\nonumber \\ & + & 
   \frac{f_{\rho'}(Q^2) g_{{\rho'} \pi\pi}(Q^2)}{M_{\rho'}^2} 
      \mbox{BW}_{\rho'}(Q^2)
\nonumber \\ & + & 
 \cdots
\end{eqnarray}
where the $\cdots$ may indicate both higher resonances ($\rho''$, \dots)
and possible non-resonant contributions.
We use Breit-Wigner resonance factors normalized to $\mbox{BW}_X(0) =1$,
\begin{equation}
   \mbox{BW}_X(Q^2) = \frac{M_X^2}{M_X^2 - Q^2 - i \sqrt{Q^2} \Gamma_X(Q^2)}
\end{equation}
The on-shell values of the various couplings can (approximately) be
determined from experimental data,
\begin{eqnarray}
  |g_{\rho\pi\pi}(M_\rho^2)| & = & 6.05
\nonumber \\
  |g_{\rho'\pi\pi}(M_{\rho'}^2) |& = &1.39
\nonumber \\
  |f_\rho(M_\rho^2) | & = & 0.117\, {\rm GeV}^2
\nonumber \\
  |f_{\rho'}(M_{\rho'})| & = & 0.18\, {\rm GeV}^2
\end{eqnarray}
If we {\em assume} that the meson couplings are approximately constant
from $Q^2 = 0$ up to the relevant resonance masses, we find
\begin{eqnarray} \label{eqnfv}
F_V(Q^2) & = & 
   \underbrace{\frac{f_\rho(M_\rho^2) g_{\rho\pi\pi}(M_\rho^2)}
   {M_\rho^2}}_{1.20} \mbox{BW}_\rho(Q^2)
\nonumber \\ & + &
   \underbrace{\frac{f_{\rho'}(M_{\rho'}^2) g_{{\rho'}\pi\pi}(M_{\rho'}^2)}
   {M_{\rho'}^2}}_{-0.12} \mbox{BW}_{\rho'}(Q^2)
\nonumber \\ & + &
   \cdots
\end{eqnarray}
Here we have used the experimental knowledge about the negative
relative phase between the $\rho$ and the $\rho'$ contributions.
Note that no reliable error estimates are possible here. For the value
${f_{\rho'}(M_{\rho'}^2) g_{{\rho'}\pi\pi}(M_{\rho'}^2)}/
   {M_{\rho'}^2} = 0.12$
we get an error of $\pm 0.02$ from the uncertainty in $M_{\rho'}$, 
but the uncertainty from $g_{\rho' \pi \pi}$ is unknown and might potentially
be larger.

We will now compare these results with the CN-VMD model. If we dominate
with the $\rho$ only, we get
\begin{equation}
  F_V(Q^2) = 1 \times \mbox{BW}_{\rho}(Q^2)
\end{equation}
Compared to (\ref{eqnfv}), we find a $20\%$ discrepancy in the strength
of the $\rho$ contribution. This has already been observed in
\cite{braaten}. 
If, however, we include the first two resonances, we get
\begin{eqnarray}
  F_V(Q^2) 
& = & \underbrace{\frac{1}{1 + \beta}}_{1.17} \times \mbox{BW}_{\rho}(Q^2)
\nonumber \\
& + & \underbrace{\frac{\beta}{1 + \beta}}_{-0.17} \times 
\mbox{BW}_{\rho'}(Q^2)
\end{eqnarray}
where we have taken $\beta$ from the fit $N=1$ in \cite{Kue92},
and with three resonances, we get (fit $N=2$ in \cite{Kue92})
\begin{eqnarray}
  F_V(Q^2) 
& = & \underbrace{\frac{1}{1 + \beta + \gamma}}_{1.16} \times 
\mbox{BW}_{\rho}(Q^2)
\nonumber \\
& + & \underbrace{\frac{\beta}{1 + \beta + \gamma}}_{-0.12} 
   \times \mbox{BW}_{\rho'}(Q^2)
\nonumber \\
& + & \underbrace{\frac{\gamma}{1 + \beta + \gamma}}_{-0.04} 
   \times \mbox{BW}_{\rho''}(Q^2)
\end{eqnarray}
These numbers show that the CN-VMD model compares reasonably well with
(\ref{eqnfv}), if the first two or three $\rho$ like states are included.
This gives us some confidence that the CN-VMD model is indeed a good
approximation. A more detailed study is in preparation \cite{Fin97}.

{\bf 2.} {$K \pi$ final states.}
There are two such final states, $\overline{K}^0 \pi^-$ and 
$K^- \pi^0$, which are
related by isospin. Due to $m_s > m_u, m_d$, both $F_V$ and $F_S$ 
may contribute \cite{truong}. Relevant resonances with
the correct quantum numbers are:
$$
\begin{array}{ll}
  J^P = 1^- \,\,(F_V): & K^*(892) \quad \mbox{and} \quad K^*(1410)
\\
  J^P = 0^+ \,\, (F_S): & K^*_0(1430)
\end{array}
$$
In \cite{fm95}, we have constructed a model with includes both the
$K^*(1410)$  and the $K_0^*(1430)$.

The relative strength $\beta_{K^*} \approx -0.135$ of the $K^*(1410)$ 
in the vector channel was
estimated
from $\mbox{BR}(K\pi)$ (note that $\beta_{K^*} \approx \beta_{\rho}$,
which is consistent with $SU(3)$ flavor symmetry expectations),
and the strength of the $K_0^*(1430)$ by matching to $O(p^4)$ 
Chiral Perturbation Theory. 
We found that the $K^*(1410)$ and the $K^*_0(1430)$
contribute with comparable strength (both about 5\% to the rate).
Thus one has to disentangle them using angular distributions, i.e.\ 
by measuring $W_B$, $W_{SA}$ and $W_{SF}$.

{\bf 3.} {$3\pi$ final states.}
Some important tasks in these final states are to
establish the presence of a scalar contribution and to measure its size,
and to put upper limits on the $\rho'$ contribution in two pion subresonances.

{\bf 4.} {$K\pi K$ and $\pi K \pi$ final states.} 
As opposed to the case of the $3\pi$, here both the vector and the 
axial vector currents can contribute (i.e.\ both the $a_1$ and the
$\rho'$, or both the $K_1$ and the ${K^*}'$). 
Theoretical predictions
for their ratio $V : A$ vary considerably 
\cite{Gom90,braaten,Dec93,Fin96,Li96}.
Thus it should be measured, using angular distributions. 
Understanding these decay modes may turn out to be very useful for 
measurements of possible non standard model CP-violation 
\cite{cp}.
This is because suitable interference terms between the vector and the
axial vector contribution can be measured without polarized taus, and
without knowledge of the $\tau$ rest frame; and because the two 
interfering amplitudes (vector and axial vector) may have comparable
strengths.

Another interesting issue regarding the $\pi K \pi$ final states are
the two axial resonances with strangeness, the $K_1(1270)$ and the
$K_1(1400)$, which both can contribute here.

We believe that there are good reasons to suspect that the $K_1$ widths
given in the particle data book \cite{RPP96} might be considerably too 
small \cite{erwin}. These reasons are: 
(i) Otherwise the $BR(\pi K \pi)$ within the CN-VMD model turn out
to be about a factor of $2 \cdots 3$ too large. Of course, the obvious
thought is that this might be due to a failure of the CN-VMD model.
However, due to the fact that most of the other predictions of the
model for three meson final states agree reasonably well
with data, and due to the
various symmetry relations between the different three meson final states,
we are actually unable to find any other natural explanation.
This led us to suspect that indeed the true $K_1$ widths might be larger.
(ii) Given that $\Gamma_{a_1} \approx 400 \cdots 600 \, \mbox{MeV}$,
the $K_1$ widths in \cite{RPP96} appear unnaturally small.
(iii) Experiments on which the numbers quoted in \cite{RPP96} are
based, have made the same
assumptions for the parametrization of background amplitudes, which
in the past
yielded very small values for $\Gamma_{a_1}$,
and which subsequently were
superseded by cleaner measurements of $\Gamma_{a_1}$ using $\tau$ decays.
\begin{figure}           
\caption{Structure functions for $\tau\to K^- \pi^- \pi^+$:
$w_A(Q^2)$ (solid) and $w_B(Q^2)$ (dashed). 
The larger solid curve with two clearly distinguishable peaks is
obtained using the RPP-96 values for $\Gamma_{K_1(1270)}$ 
and $\Gamma_{K_1(1400)}$, the smaller solid curve using
$\Gamma_{K_1(1270)} = \Gamma_{K_1(1400)} = 250\, \mbox{MeV}$}
\label{figk1}
\mbox{\begin{sideways}{\epsfig{figure=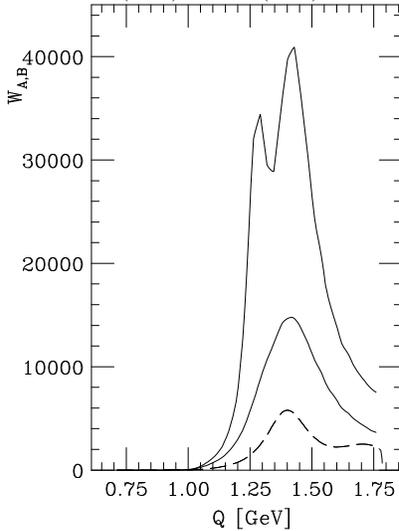,width=7.cm}}
 \end{sideways}}                %
\end{figure}
Thus our conclusion is that one should {\em measure} the $K_1$ resonance
parameters in $\tau \to \pi K \pi \nu_\tau$. In Fig.~\ref{figk1} we
show that indeed $w_B$ is very sensitive to the $K_1$ parameters.
These issues will be discussed in more detail in a forthcoming
publication \cite{Fin97}.

\section{Conclusions}
Let us conclude by making some suggestions to experimentalists about
what we believe to be interesting and feasible measurements.

First, let us consider decays with three pions. We would suggest to
measure $w_D$ and $w_E$ for both charge modes, $2 \pi^- \pi^+$ and
$2 \pi^0 \pi^-$. These two structure functions might very well be 
different for the two modes. This would be the first evidence for the
[300] partition class, which decays differently into $2\pi^-\pi^+$ and
$2\pi^0 \pi^-$.
Furthermore, we would like to urge to measure $w_D$ and $w_E$ for
$Q^2$ less or nearly equal to $0.8\,\mbox{GeV}^2$ for the $2\pi^-\pi^+$
charge mode. There is most probably a zero and a change of sign both
in $w_D$ and $w_E$ in this mode, somewhere between $Q^2 = 9 M_\pi^2 \cdots
0.8\, \mbox{GeV}^2$.

We also suggest to measure the scalar contribution via $w_{SB}$, 
and to try to put limits on the $\rho'$ contribution in the two-pion 
subresonances.

Secondly, regarding $\tau \to \pi K \pi \nu_\tau$, we suggest to 
measure the parameters of the $K_1$ resonances using this $\tau$ decay
mode.

\section*{Acknowledgments}
%
We thank J. Gasser and J.H. K\"uhn for very fruitful discussions. 

	This work is supported in part by the National Science
	Foundation (Grant \#PHY-9218167), by HCM, EEC-Contract No.
	CHRX-CT920026 (EURODA$\Phi$NE), by BBW, Contract No. 93.0341, by the
	Deutsche Forschungsgemeinschaft, 
	and by Schweizerischer Nationalfonds.

%

\clearpage

\end{document}